\documentclass[twocolumn]{revtex4-2} 
\usepackage{graphicx}
\usepackage{amssymb}
\usepackage{amsmath}
\usepackage{color}
\usepackage[normalem]{ulem}

\newcommand{\eV}{\mathrm{eV}}

\begin{document}
\title{Constrains of the axion-like particle from black hole spin superradiance}
\author{Lei-dong Cheng}
\email{chengleidong@mail.sdu.edu.cn}
\author{Hong Zhang}
\email{hong.zhang@sdu.edu.cn}
\author{Shou-shan Bao}
\email{ssbao@sdu.edu.cn}
\affiliation{Institute of Frontier and Interdisciplinary Science,\\
Key Laboratory of Particle Physics and Particle Irradiation, Shandong University, Qingdao 266237, P. R. China}
\begin{abstract}%
The mass of the axion-like particles could be constrained by the observed black hole spin distribution. In this work, we update the previous calculations using the recently improved superradiance formula, which is much more accurate. The effect of the merger time scale is also carefully investigated with Bayesian analysis. After integration of the merger time $\tau_\mathrm{M}$, we find two favoured mass ranges $7.94\times10^{-13}~\eV\leq\mu\leq8.71\times10^{-13}~\eV$ and $1.44\times10^{-12}~\eV\leq\mu\leq1.74\times10^{-12}~\eV$ . These two favored ranges do not depend on the prior distribution of the $\tau_\mathrm{M}$. We also find the strength of evidence for the range $-12.80\leq\log_{10}(\mu/\eV)\leq-12.34$, which is excluded in the analysis with fixed $\tau_\text{M}$, could increase by several orders of magnitude with $\tau_\mathrm{M}$ averaged.
\end{abstract}
\maketitle
\section{Introduction}

Ultralight particles can form gravitational bound states around a spinning black hole (BH) if the Compton wavelengths are comparable to the BH horizon size. Especially, the bound-state eigenfrequency can have a positive imaginary part with proper choices of the host BH's mass and spin, as well as the ultralight particle mass. This is often called superradiance in literatures \cite{Penrose:1969pc,Misner:1972kx}. The growth of the bound states extracts energy and angular momentum continuously from the host BH until the BH spin is below some critical value or nonlinear effects become important \cite{Arvanitaki:2010sy,Yoshino:2012kn,Baryakhtar:2020gao}. In this process, the BH can lose a significant portion of the initial angular momentum. Experimentally, the gravitational wave (GW) events measured by the Laser Interferometry Gravitational-Wave Observatory (LIGO)~\cite{LIGOScientific:2014pky},  Virgo~\cite{VIRGO:2014yos} and KAGRA~\cite{KAGRA:2020tym} (LVK) can tell us the spins of the two BHs before merging. The obtained BH spin distribution from many binary black hole (BBH) events provides us with a valuable tool to probe the ultralight particles~\cite{Stott:2018opm,Arvanitaki:2016qwi,Cardoso:2018tly,Brito:2017zvb,Baumann:2019ztm,Ng:2019jsx,Payne:2021ahy,Khodadi:2021gbc,Yuan:2021ebu}.

In this work, we focus on the superradiant effect of axion-like-particles (ALPs), which is one of the most popular candidates of cold dark matter~\cite{Hu:2000ke,Bertone:2004pz,Arvanitaki:2010sy, Hui:2021tkt,Oks:2021hef,Cicoli:2021gss}. The available LIGO data are sensitive to the ALP mass in the range from $10^{-13.5}$~eV to $10^{-11}$~eV \cite{Brito:2017zvb}. Constrains on even lighter ALP requires observations with supermassive BHs \cite{Chung:2021roh,Zu:2020whs,Davoudiasl:2019nlo}. The calculation of the constraint depends on the eigenfrequencies of the boson clouds.   Previous studies use the analytic approximation given by Detweiler in Ref.~\cite{Detweiler:1980uk}, which does not agree with the numerical calculation in Ref.~\cite{Dolan:2007mj}. Recently, a mistake in this approximation is pointed out and an improved expression is provided in Ref.~\cite{Zhang:2022}. The improved approximation has a compact form and agrees very well with the numerical calculation.

In this work, we update the previous constraint of the ALP mass from the BH spin distribution with the improved approximation. We also improve the previous analysis in the following two aspects. In Ref.~\cite{Ng:2019jsx} the authors made a hierarchy Bayesian analysis, and in Ref.~\cite{Ng:2020ruv} they analyzed 45 BBHs events in GWTC-1 and GWTC-2 with a distribution model $\chi^\alpha (1-\chi)^\beta$ for the prior distribution of the black spins at formation, where $\alpha, \, \beta$ are free parameters in addition to $\mu$ in their analysis. In Ref.~\cite{Fernandez:2019qbj}, they analyzed 10 BBHs events with two different prior for initial spin distribution. In all these previous studies, the merger time scale $\tau_\text{M}$ is fixed. In this work, we made an analysis with all BBHs in three phases of GTWC data reported by LVK collaboration~\cite{LIGOScientific:2018mvr,LIGOScientific:2020ibl,LIGOScientific:2021djp}, only excluding the events with neutron stars (GW170817, GW190425, GW190426\_152155, GW190814, GW191219\_163120,  GW200115\_042309,  GW200210\_092254). We consider three different initial spin prior distributions to identify the effects of the prior distribution.  Since the calculated spin distribution of BH strongly depends on the time scale $\tau_\text{M}$, we also take into account the distribution of $\tau_\mathrm{M}$ in this work.  

This paper is organized as follows. In Sec.~\ref{sec:superradiance} we give a brief overview of the superradiance and the Regge trajectories. In Sec.~\ref{sec:results}, we present the Bayesian analysis and our results using the LIGO data. A short summary is presented in Sec.~\ref{sec:summary}.

\section{Superradiance and Regge Plots}\label{sec:superradiance}
The spacetime outside the horizon of a spinning BH can be described by the Kerr metric with the Boyer-Lindquist coordinates  $x^\mu=(t,r,\theta,\phi)$,
\begin{align}
\begin{split}
ds^{2}&=\left(1-{\frac {2 r_g r}{\Sigma }}\right) dt^{2}+{\frac {4 r_g r a\sin ^{2}\theta }{\Sigma }}\,dt\,d\phi
-{\frac {\Sigma }{\Delta }}dr^{2}\\
&-\Sigma d\theta ^{2}
-\left(r^{2}+a^{2}+{\frac {2 r_g r a^{2}}{\Sigma }}\sin ^{2}\theta \right)\sin ^{2}\theta \ d\phi ^{2},
\end{split}
\end{align}
where,
\begin{subequations}
\begin{align} 
r_{g} &= {GM},\qquad a={\frac {J}{M}},\\ 
\Sigma &=r^{2}+a^{2}\cos ^{2}\theta,\\
r_\pm &=r_g\pm\sqrt{r_g^2-a^2},\\
\Delta &=r^{2}-2r_g r+a^{2}=(r-r_+)(r-r_-),
\end{align}
\end{subequations}
with $M$ and $J$ the mass and angular momentum of the BH, respectively. The outer event horizon $r=r_+$ is obtained by setting $\Delta$ to be zero, which also requires $a\leq r_g$. A dimensionless spin parameter is defined as $\chi=a/r_g\leq 1$.  The stationary limit surface is determined by $g_{00} = 0$. The region bounded by the outer event horizon and the stationary limit surface is called the ergosphere. Interesting superradiant phenomena happen inside the ergosphere. We refer the interested readers to a recent review in Ref.~\cite{Brito:2015oca} for details.

Massive scalar boson can form bound states around a Kerr BH. These bound states, also called boson clouds, are described by the Klein-Gordon equation,
\begin{align}
\left(g^{ab}\nabla_a\nabla_b-\mu^2\right)\Phi=0,
\end{align}
where $g^{ab}$ is the inverse of the Kerr metric, $\nabla_a$ is the covariant derivative, and $\mu$ is the boson mass. The solution of this equation can be written with separation of variables \cite{Brill:1972xj},
\begin{align}
\Phi=\sum_{nlm}e^{-i\omega_{nlm} t} e^{im\phi} S_{lm}(\theta)R_{nlm}(r) +\text{c.c.,}
\end{align}
where $n\geq 0$ is the principal number, $l$ and $m$ are the orbital angular momentum and the magnetic number, respectively. The properties of the spheroidal harmonic functions $S_{lm}(\theta)$ are explained in Ref.~\cite{Berti:2005gp}. Due to the superradiant effect, the eigenfrequency $\omega$ has an imaginary part. In this work, we calculate $\omega$ with the recently improved analytic approximation in Ref.~\cite{Zhang:2022}, which has a compact form and agrees with the numerical calculation.

The eigenfrequency $\omega$ depends only on $r_g\mu$ and $\chi$. With proper values of these two parameters, the imaginary part of $\omega$ is positive, causing the amplitude of the bound state to increase exponentially with time. The number of bosons, which is proportional to the square of the wavefunction amplitude, also increases exponentially,
\begin{align}\label{eq:N}
N_{nlm}(t)=N_{nlm} (0) \exp\left(\Gamma_{nlm} t\right),
\end{align}
where $\Gamma_{nlm}=2\,\text{Im}(\omega_{nlm})$ is also a function of $r_g\mu$ and $\chi$. The growth of the boson cloud is balanced by decreases of the mass and the angular momentum deposited initially in the host BH. In this process, the BH could lose at most $10\%$ of its mass \cite{Brito:2017zvb}, which is much smaller than the uncertainty of the available data from the GW telescopes. Hence we ignore the change of the BH mass in this work. On the other hand, the superradiance is very efficient in extracting the angular momentum from the host BH. For fixed $r_g\mu$ and merger time scales $\tau_\mathrm{M}$, the BH with e-folding number $\Gamma_{nlm}\tau_\mathrm{M} > 180$ cannot exist \cite{Fernandez:2019qbj}. For fixed $\mu$ and $\tau_\mathrm{M}$, this relation gives an exclusion region on the $M-\chi$ plane in which no BHs should be observed. The boundary of this region is often called the Regge trajectory in literatures.

\begin{figure}
\centering
\includegraphics[clip, trim=0.35cm 0cm 0cm 0cm, width=0.44\textwidth]{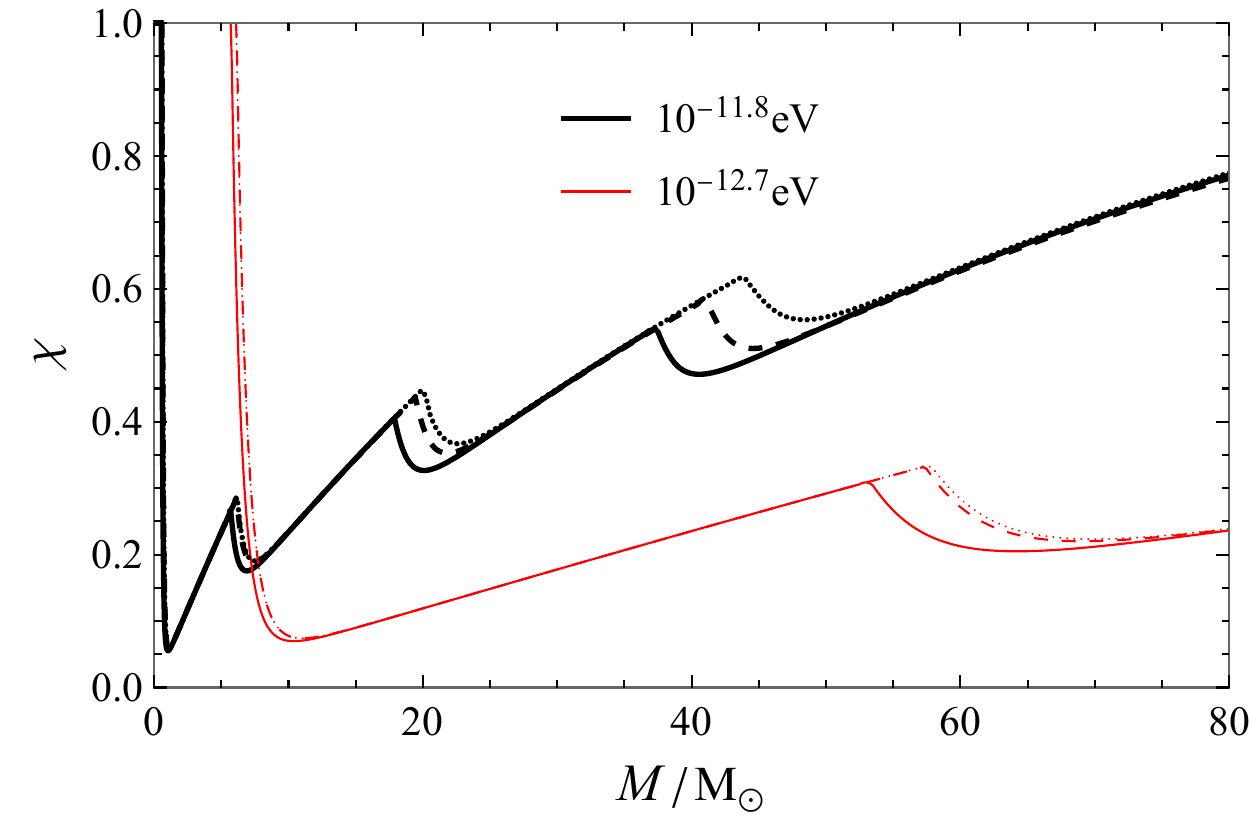}
\caption{The Regge trajectories of superradiance for two different boson masses $10^{-11.8}~\eV$ (black curves) and $10^{-12.7}~\eV$ (red curves), with $\tau_\mathrm{M}=10^{10}$~yr. The dotted curves are from the previous analytic approximation in Ref.~\cite{Detweiler:1980uk}. The dashed curves are from the recently improved analytic approximation in Ref.~\cite{Zhang:2022}, with $n=0$ only. The solid curves are from the recently improved analytic approximation with $n$ summed from 0 to 8.\label{fg:regge}}
\end{figure}

Eq.~\eqref{eq:N} includes only the superradiant growth. If other astrophysical processes such as accretion are taken into account, the Regge trajectory turns out to be an attractor  \cite{Brito:2014wla}. As a result, a BH produced in the exclusion region is expected to end up on the Regge trajectory with roughly the same mass, while the BHs initially outside of the exclusion region will not be affected significantly. In Fig.~\ref{fg:regge}, we show the  Regge trajectories for two typical boson masses. The region above each curve is the corresponding exclusion region. The curves with $\mu=10^{-12.7}$~eV have a much larger exclusion region compared to the curves with  $\mu=10^{-11.8}$~eV. If many BHs are observed to have $\chi$ close to $0.3$, then the scalar boson with a mass close to $10^{-12.7}$~eV are unfavored with high confidence.

\begin{figure}
\centering
\includegraphics[scale=0.6]{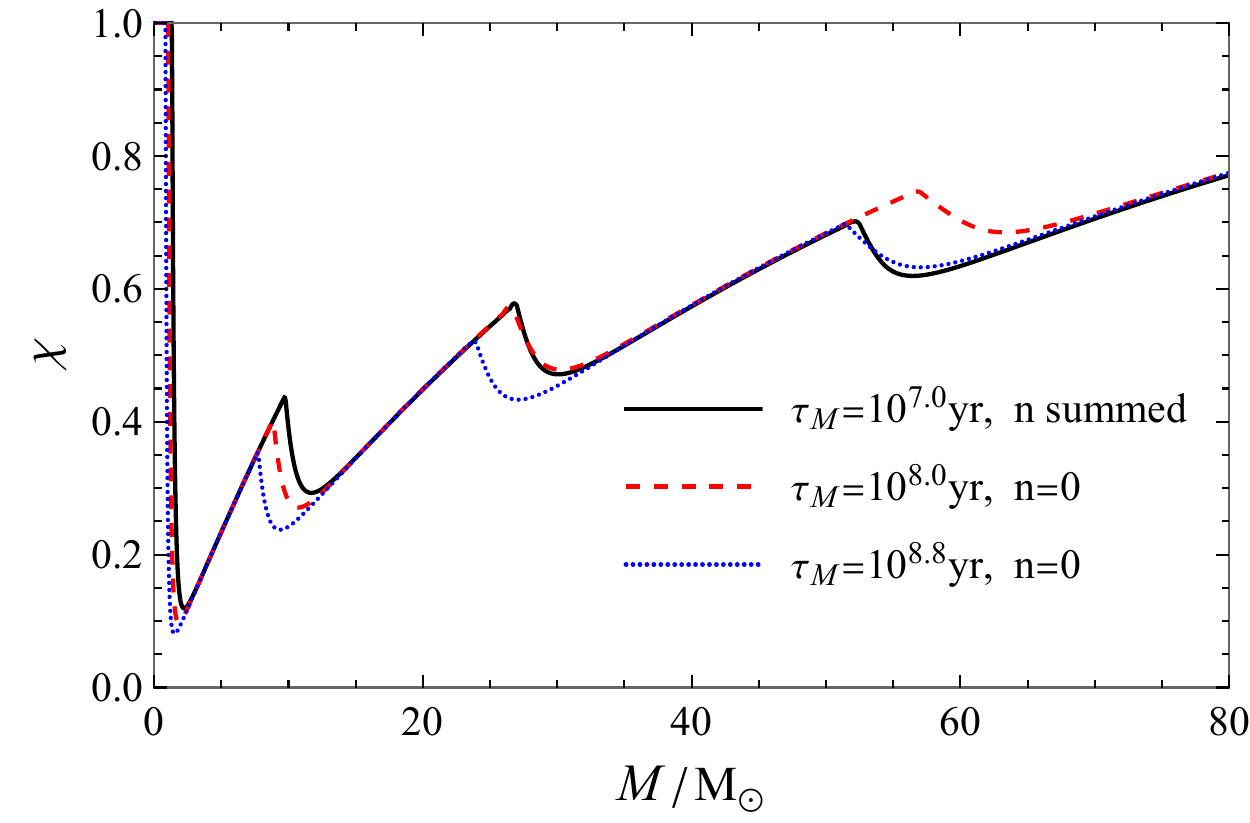}
\caption{The change of the Regge trajectories by varying $\tau_\mathrm{M}$. The value of $\mu$ is fixed as $10^{-11.8}$~eV for all three curves. The solid curve is from the improved analytic approximation with $n$ summed to 8 and $\tau_\mathrm{M}=10^{7.0}$~yr. The dashed and dotted curves are from the previous approximation with $\tau_\mathrm{M}=10^{8.0}$~yr and $10^{8.8}$~yr, respectively. 
\label{fg:vary_tau}}
\end{figure}

In Fig.~\ref{fg:regge}, we also compare the recently improved analytic approximation in Ref.~\cite{Zhang:2022} to the previous one in Ref.~\cite{Detweiler:1980uk}. Especially, we summed the contribution of $n$ from 0 to 8, while previous calculations of the Regge trajectories only consider the contribution of $n=0$. The three curves with fixed boson mass $\mu$ differ significantly close to the wiggles. One may expect the difference can be absorbed by varying $\mu$ and/or $\tau_\mathrm{M}$. Since $\Gamma_{nlm}$ depends only on $r_g\mu$ and $\chi$, changing the value of $\mu$ is equivalent to a rescaling of the BH mass $M$, which squeezes or stretches the whole curve horizontally. Thus the difference of the curves cannot be eliminated by changing $\mu$. In Fig.~\ref{fg:vary_tau}, we show the effect of varying $\tau_\mathrm{M}$. We find although one could merge the curves at a single wiggle, it is impossible to remove the differences at all wiggles. Therefore, we conclude that the difference between the two analytic approximations cannot be absorbed in the parameters. With more data available in the future, the improved approximation may give a significantly different constraint of $\mu$.

\section{Bayesian Analysis and Results}\label{sec:results}
In this work, we use the hierarchical Bayesian analysis method, which is explained in detail in Ref.~\cite{Ng:2019jsx}. The hyper-posterior distribution $p(\mu\vert \{d^i\})$ for the light boson mass $\mu$ with the given data set $\{ d^i\}$ is expressed as,
\begin{align}
p(\mu \vert \{d^i\})\propto p(\{d^i\}\vert \mu) \pi(\mu),
\end{align}
where the $\pi(\mu)$ is the hyper-prior distribution of the mass $\mu$. The data from LIGO-Virgo-KAGRA can be used to constrain the scalar mass within the range $10^{-13.5} ~\eV\leq\mu\leq 10^{-11}~\eV$ \cite{Brito:2017zvb}. Without any prior knowledge, we employ a log-uniform distribution in this range. We further assume the observations of GWs events are independent of each other. Then one could write,
\begin{align}\label{eq:master-1}
p(\{d^i\}\vert \mu)=\prod_i p(d^i\vert \mu)=\prod_i \int p(d^i \vert \theta^i) p(\theta^i\vert\mu) d\theta^i,
\end{align}
where $\theta^i =(M_{1}^i,M_{2}^i, \chi_{1}^i,\chi_{2}^i,\tau_\mathrm{M}^i)$ are the  parameters of the $i\-$th observed merger event, $p(\theta^i\vert \mu)$ is the expected distribution for the event, and $p(d^i\vert\theta^i)$ is the individual likelihood of the the $i\-$th event. In this first step, we assume all the binaries have the same fixed merger time. Then the second term on the right hand side of Eq.~\eqref{eq:master-1} can be expanded as,
\begin{align}
p(\theta^i\vert \mu) d\theta^i =  \pi(M_1^i, M_2^i) \prod_{j=1}^2 p(\chi_j^i\vert M_j^i,\mu,\tau_\mathrm{M}) dM_j^i d\chi_j^i,\label{eq:master}
\end{align}
where $\pi(M_1^i,M_2^i)$ is the prior on the masses of the two BHs in the $i$th merger event. The posterior defined as above depends on the astrophysical distributions of BH spins and masses at birth. In this work, we take the BH mass prior to be a uniform distribution. And we consider the following three initial spin distribution scenarios,
\begin{align}\label{eq:pi-chi}
\pi(\chi)=\begin{cases}
1& \text{Flat spin distribution,}\\
2(1-\chi)& \text{Low spin distribution,}\\
2\chi& \text{High spin distribution.}
\end{cases}
\end{align}
The spin distribution for the BH at merger with superradiant effect can be written as \cite{Fernandez:2019qbj},
\begin{align}
p(\chi\vert M,\mu,\tau_\mathrm{M})=\pi(\chi)\theta(\chi_\mathrm{Regge}-\chi)+\alpha \delta(\chi-\chi_\mathrm{Regge}),
\end{align}
where the $\chi_\mathrm{Regge}(M,\mu,\tau_\mathrm{M})$ is the value of the Regge trajectories and $\alpha=\int^1_{\chi_\mathrm{Regge}} \pi(\chi)d\chi$. For the flat spin distribution, $\alpha$ is equal to $1-\chi_\mathrm{Regge}$. 

\begin{figure*}
\centering
\hspace{0.3cm}
\includegraphics[scale=0.65]{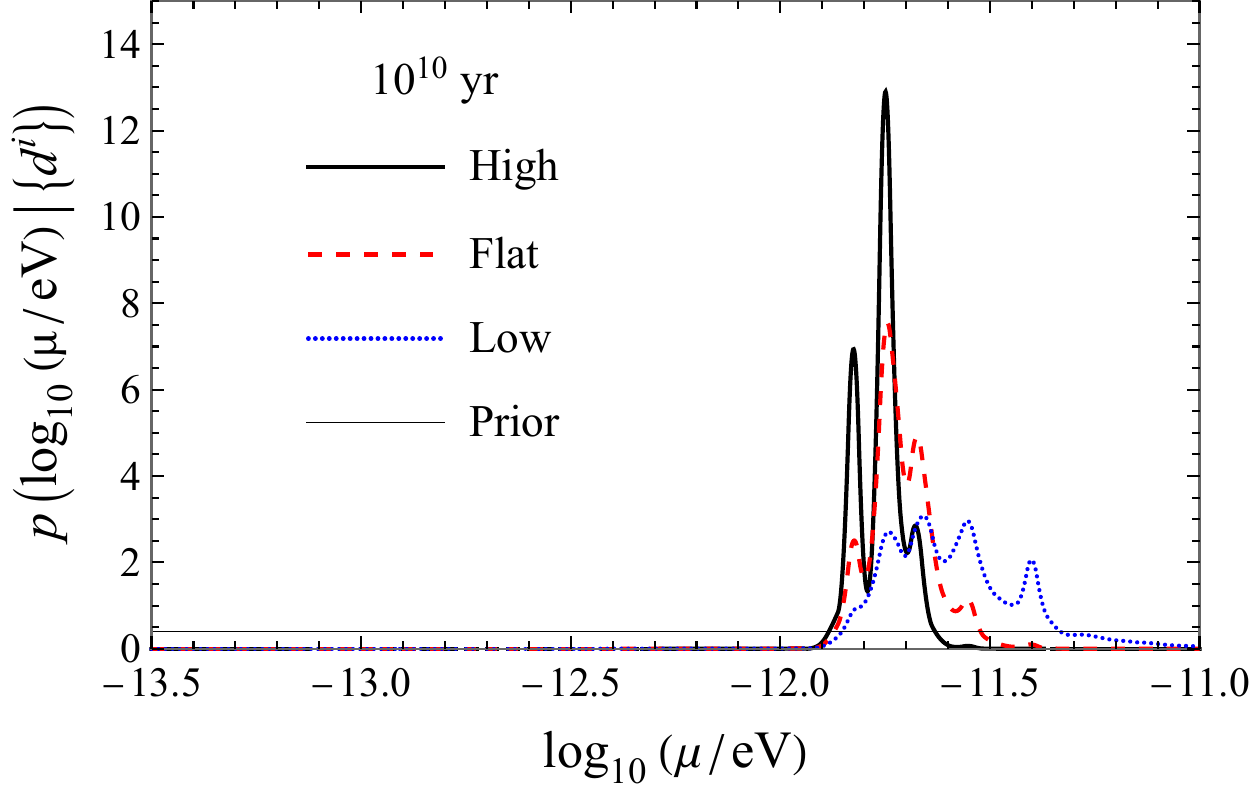}
\hspace{0.3cm}
\includegraphics[scale=0.65]{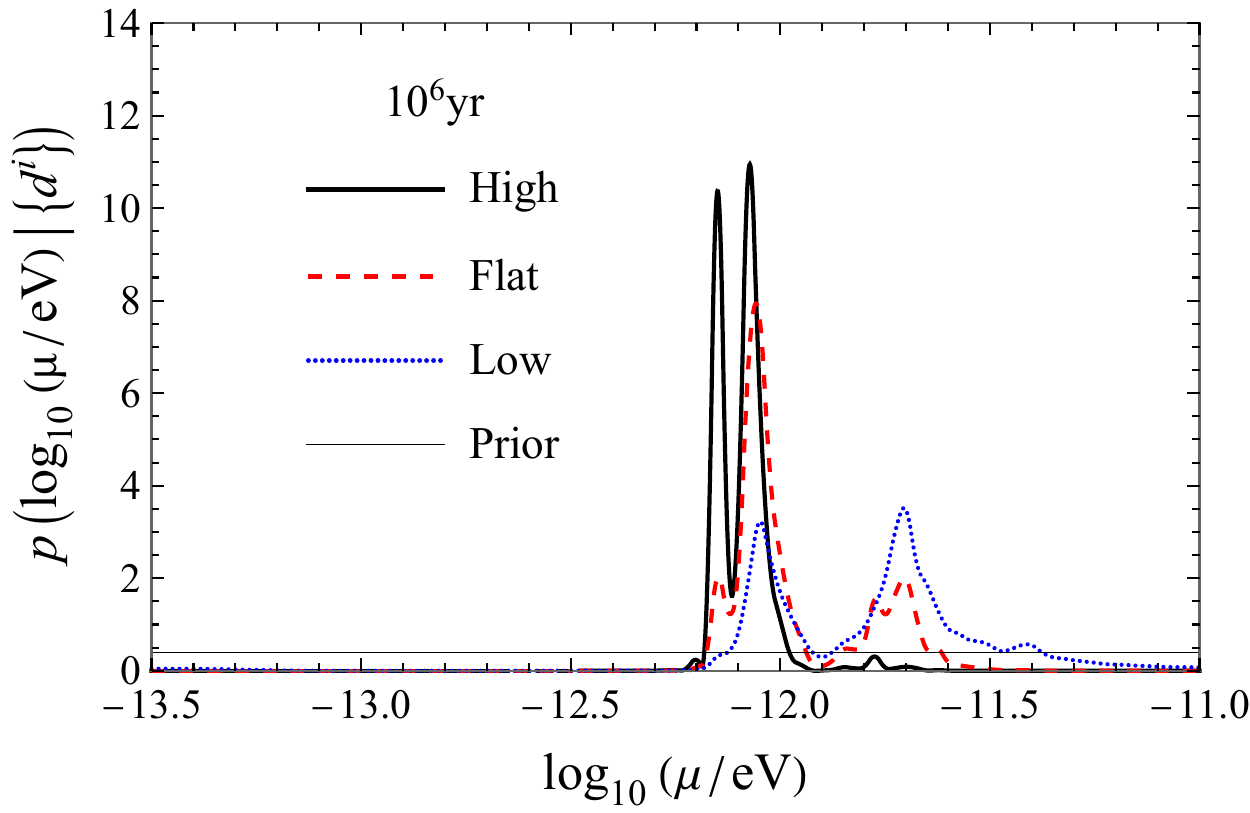}\\
\includegraphics[scale=0.68]{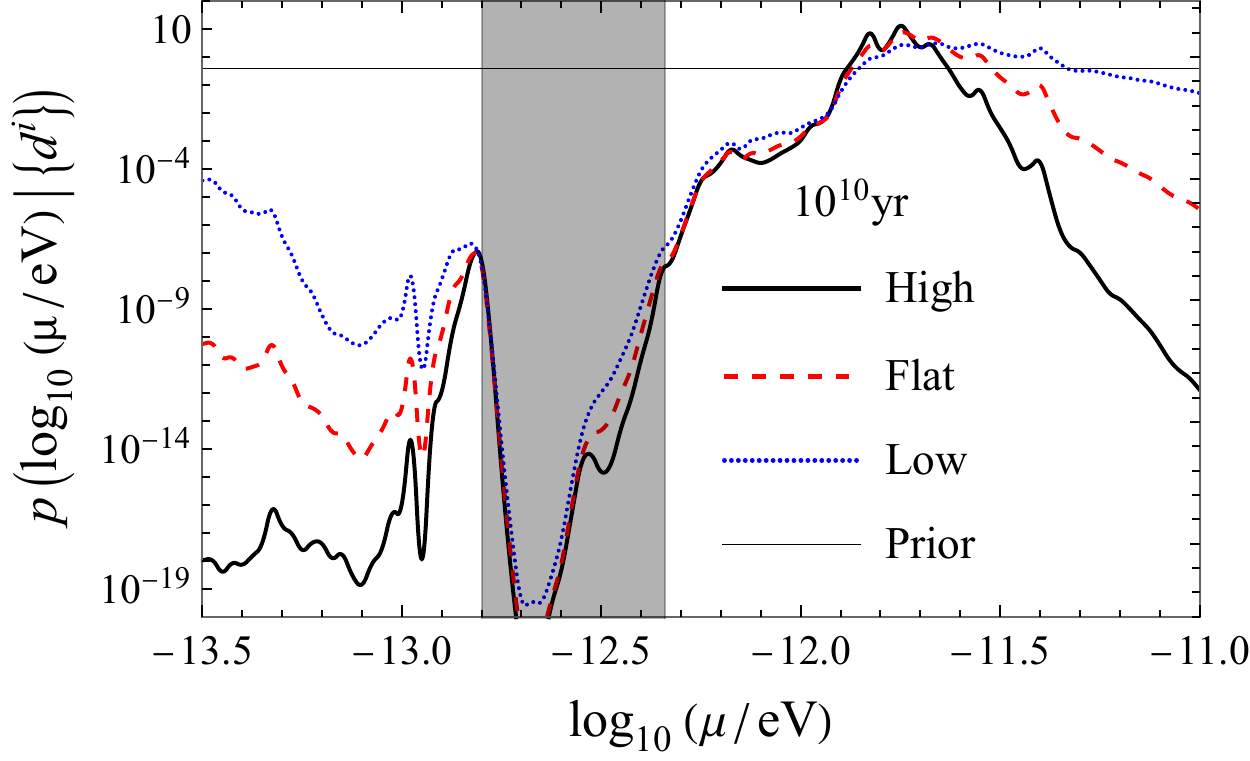}
\hspace{0.0cm}
\includegraphics[scale=0.68]{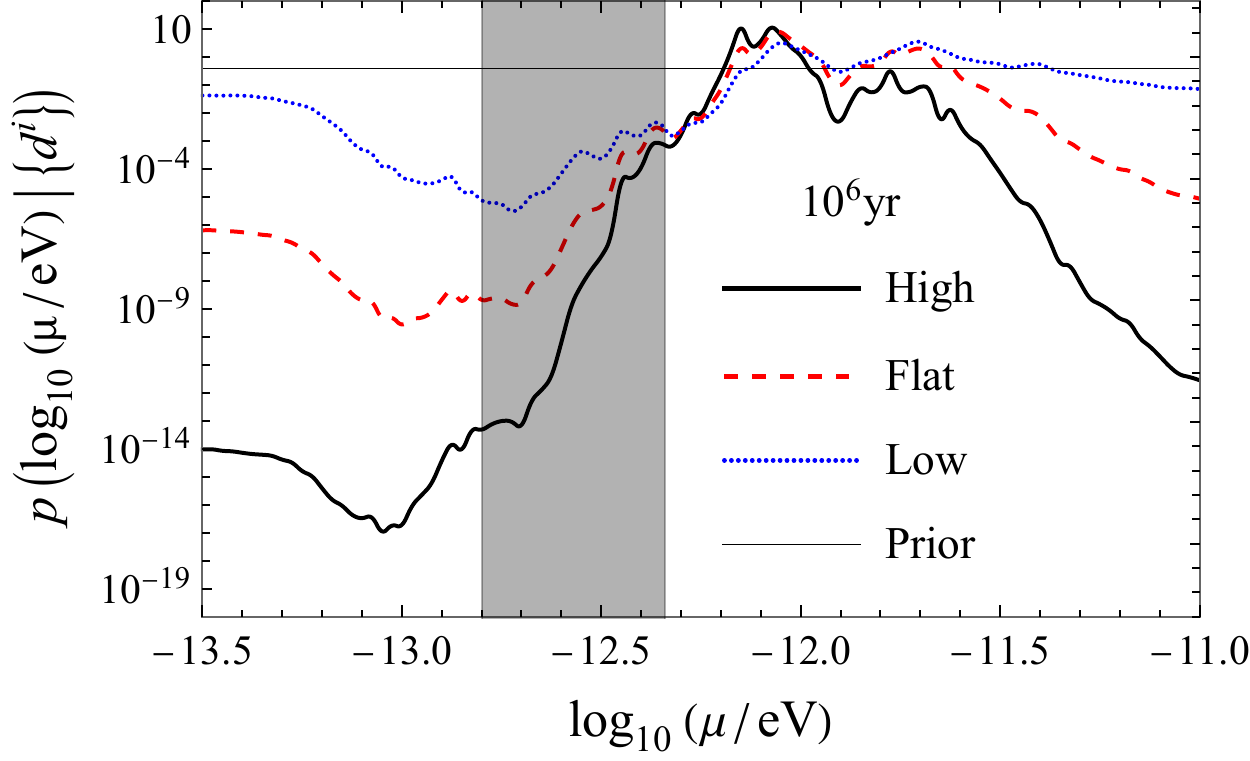}
\caption{The posterior distribution for boson mass with three different initial spin distributions and two different merger time scale $\tau_\mathrm{M}$. For each merger time, we consider three cases for the initial spin distribution given in Eq.~\eqref{eq:pi-chi}: high (black solid), flat (red dashed), and low (blue dotted). The horizontal black thin line is the log-uniform distribution for $\mu$ prior. The shadowed bands highlight the range of unfavored $\mu$ with $\tau_\mathrm{M} = 10^{10}$~yr.}\label{fg:posterior}
\end{figure*}

We perform the calculation in Eq.~(\ref{eq:master-1}) by Monte Carlo integral with LIGO posterior samples. This allows us to evaluate the likelihood without re-analyzing the original LIGO data~\cite{Wysocki:2018mpo}. The posterior distributions of the boson mass are shown in Fig.~\ref{fg:posterior} with two merger times $10^{10}$~yr and $10^6$~yr. The results depend on the prior distribution of the initial spin and the merger time quite strongly. In the high spin hyper-prior scenario, the observed data favors the boson mass range $-12.16\leq\log_{10}(\mu/\eV)\leq-12.05$ with Bayesian factor $B=16.45$ for $\tau_\mathrm{M}=10^6$~yr, and the mass range $-11.83\leq\log_{10}(\mu/\eV)\leq-11.72$ with $B=16.54$ for $\tau_\mathrm{M}=10^{10}$~yr. All the mass ranges obtained in this work are at 68\% confidence level. In the lower-left panel of Fig.~\ref{fg:posterior}, the posterior in the shadowed region is clearly small and independent on the hyper-prior of spin. It means the GWTC data strongly disfavors the existence of a light boson with a mass in this narrow range. This is consistent with the observation in Ref.~\cite{Ng:2020ruv}, in which $\tau_\mathrm{M}$ is fixed at $10^7$~yr. If choosing $\tau_\mathrm{M}=10^6$~yr, the posterior is still small, but the value depends strongly on the spin hyper-prior. Therefore, it is necessary to take into account the prior distribution of the merger time.

\begin{figure*}
\centering
\includegraphics[scale=0.6]{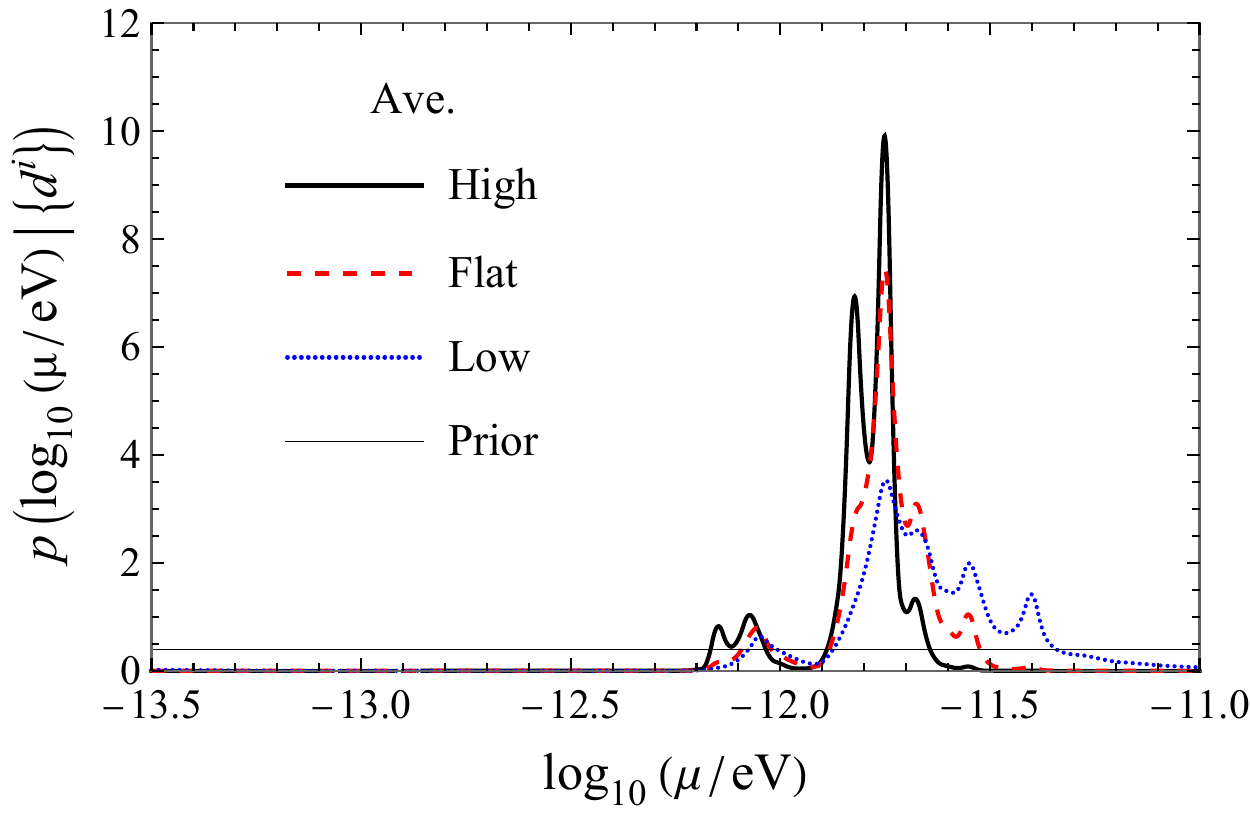}
\hspace{0.5em}
\includegraphics[scale=0.628]{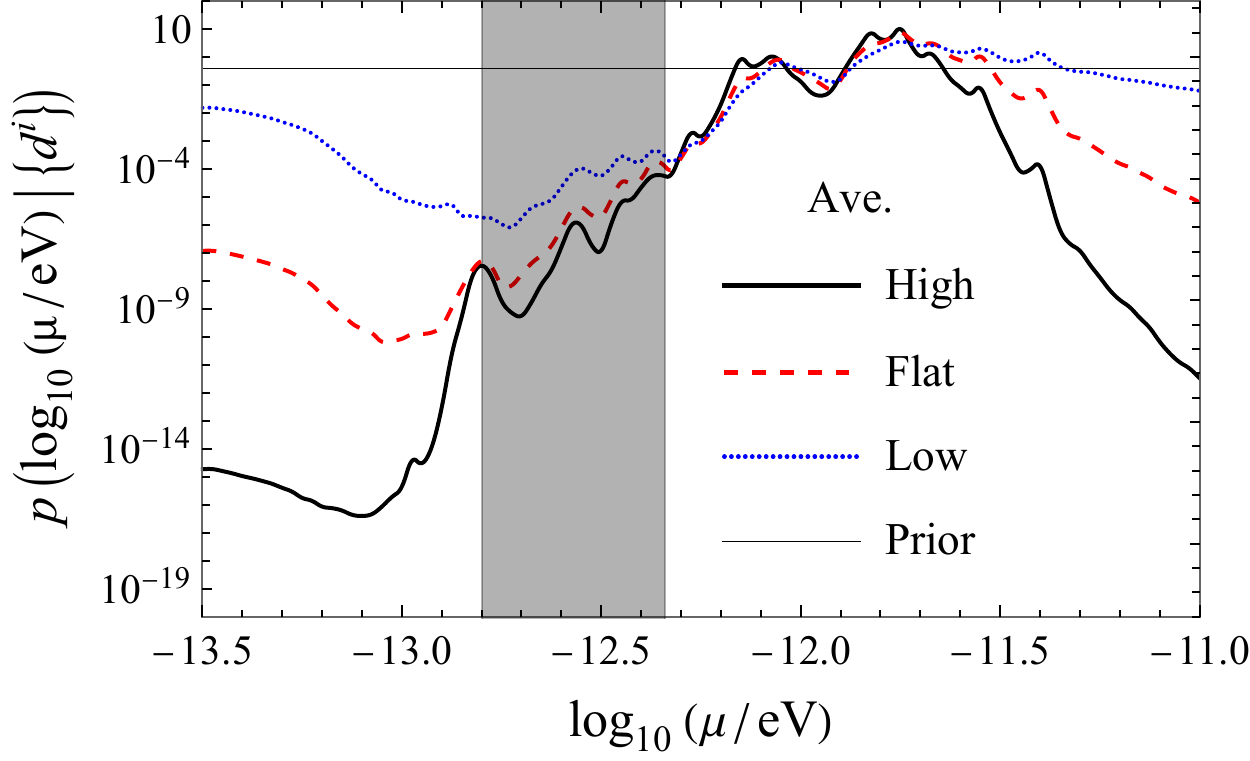}
\caption{The plot of time-averaged posterior distribution for boson mass. The three thick curves are the results with different initial spin hyper-prior distributions of the BH given in Eq.~\eqref{eq:pi-chi}: high (black solid), flat (red dashed), and low (blue dotted).  The horizontal black thin line is the log-uniform distribution for $\mu$ prior. The shadowed bands highlight the range of unfavored $\mu$ with $\tau_\mathrm{M} = 10^{10}$~yr.
}\label{fig:time_avg}
\end{figure*}

\begin{figure*}
\centering
\includegraphics[scale=0.58]{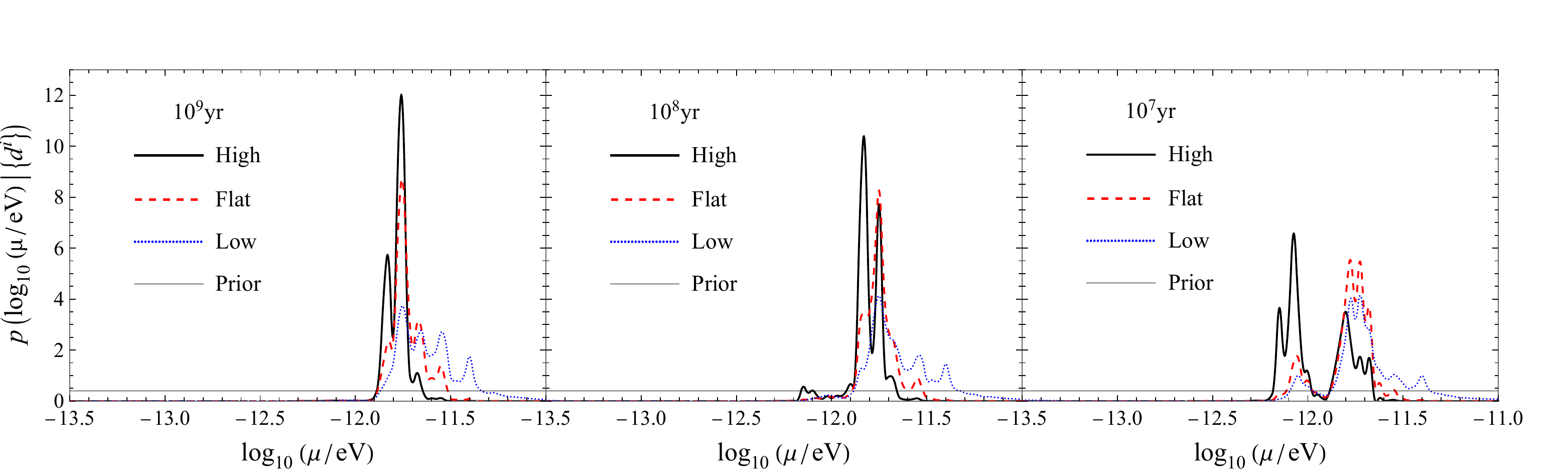}
\caption{The posteriors with different merger times $\tau_\mathrm{M}$. The three thick curves are the results with different initial spin hyper-prior distributions of the BH given in Eq.~\eqref{eq:pi-chi}: high (black solid), flat (red dashed), and low (blue dotted).  The horizontal black thin line is the log-uniform distribution for $\mu$ prior. }
\label{fig:peak_time}
\end{figure*}

\begin{figure*}
\centering
\includegraphics[scale=0.6]{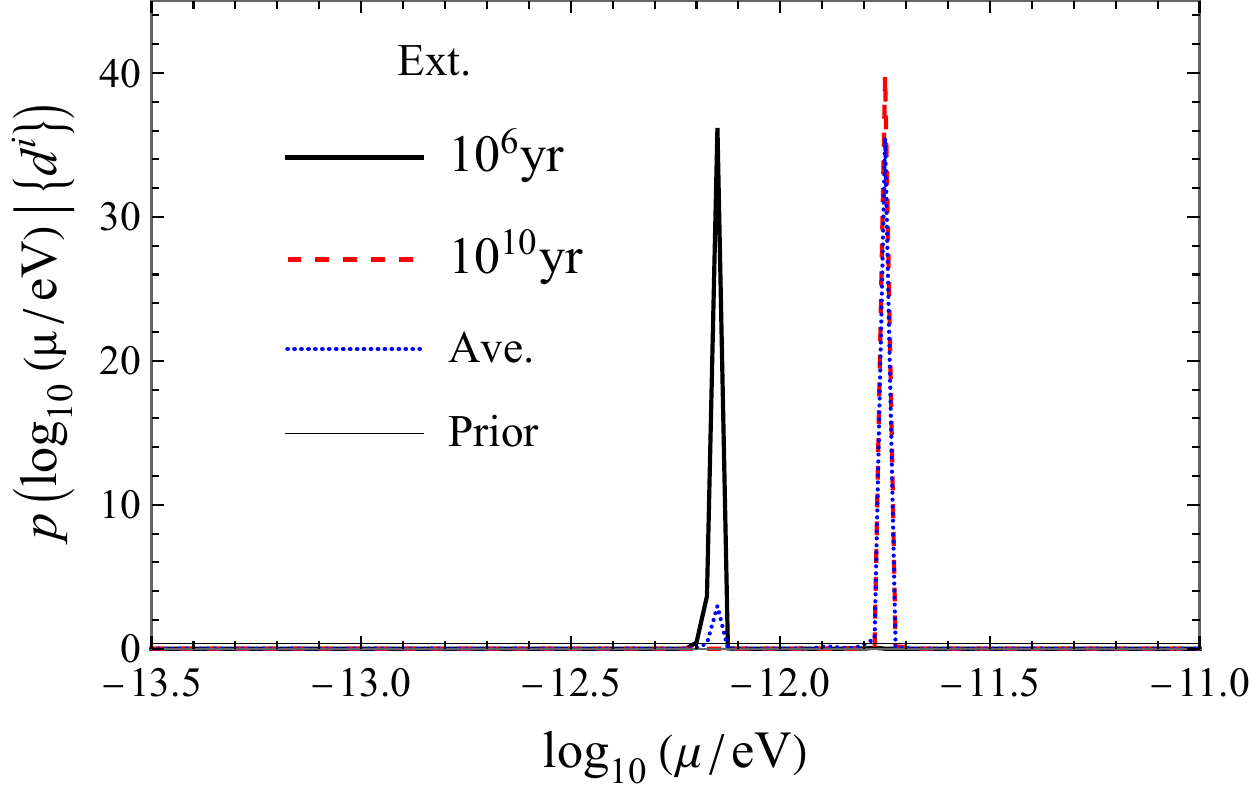}
\hspace{0.5cm}
\includegraphics[scale=0.51]{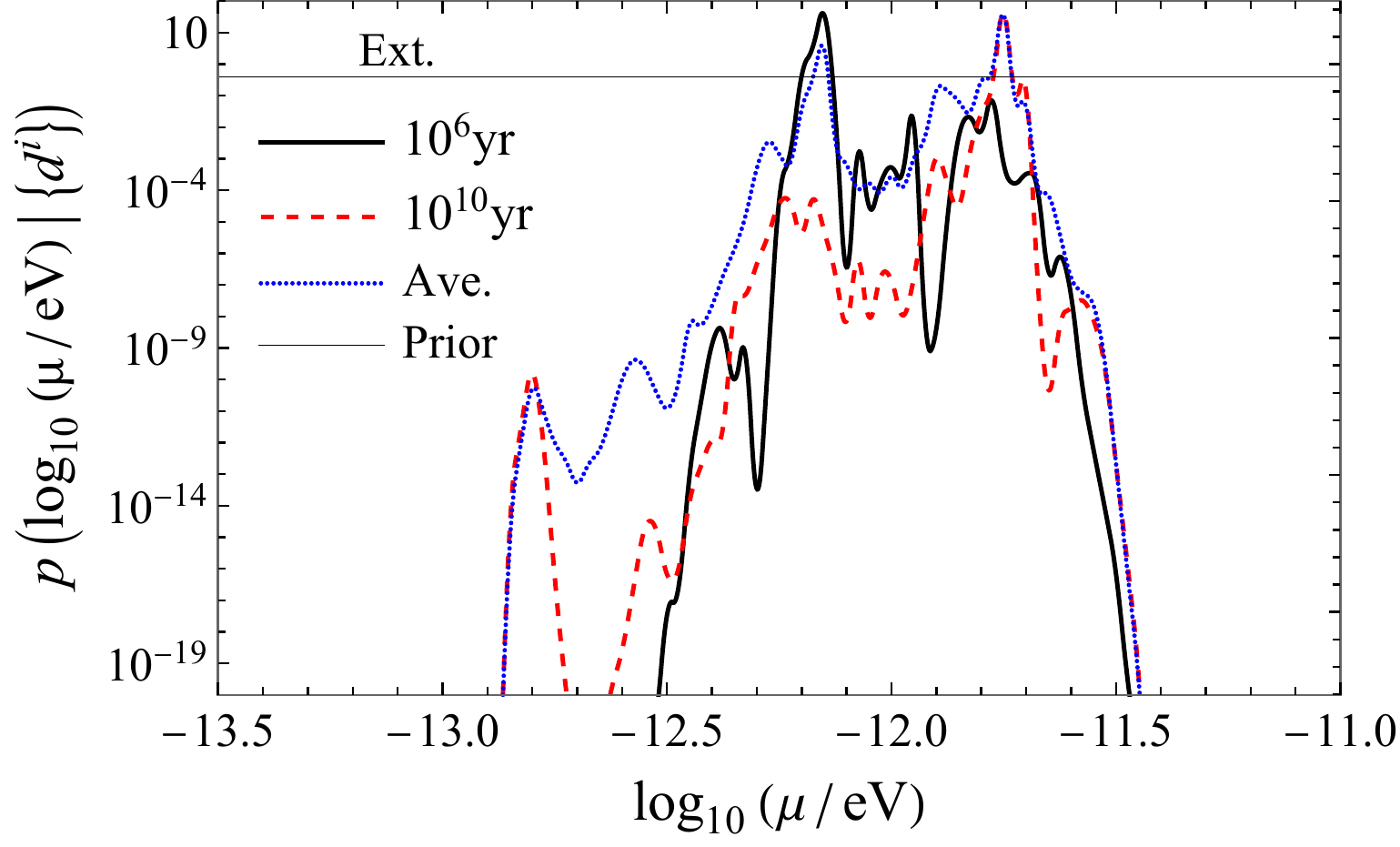}
\caption{The posterior for extreme hyper-prior of BH spin with merger time scales $\tau_\mathrm{M} = 10^6$~yr (black solid curve) and $10^{10}$~yr (red dashed curve). The posterior with $\tau_\mathrm{M}$ averaged (blue dotted curve) is also shown for comparison. See text below Eq.~\eqref{eq:master}.}
\label{fig:maximum}
\end{figure*}

To investigate the dependence of the result on the prior distribution of the merger time, we take $\tau_\mathrm{M}$ as a free parameter next. In this case, the second term on the right-hand side of Eq.~\eqref{eq:master-1} can be expanded as,
\begin{align}
\begin{split}
p(\theta^i\vert \mu) d\theta^i =&  \pi(M_1^i, M_2^i,\tau_\mathrm{M}^i) d\tau_\mathrm{M}^i \\
&\times \prod_{j=1}^2 p(\chi_j^i\vert M_j^i,\mu,\tau_\mathrm{M}^i) dM_j^i d\chi_j^i.\label{eq:master}
\end{split}
\end{align}
We assume the prior of $\tau_\mathrm{M}$ is independent on the BH masses, and is uniform in logarithmic scale in the range $10^6\sim 10^{10}$~yr, i.e. $\pi\left(\log_{10}(\tau_\mathrm{M}/\text{yr})\right)$ equals $0.25$ for $6<\log_{10}(\tau_\mathrm{M}/\text{yr})<10$ and 0 elsewhere. The marginalized posterior distribution for $\mu$ from Eq.~\eqref{eq:master-1} is a time-averaged distribution. The results are shown in Fig.~\ref{fig:time_avg}. The curves in the disfavored range with $\tau_\mathrm{M} = 10^{10}$~yr, which is shown with shadow in Fig.~\ref{fig:time_avg}, depends on the spin prior distribution. The strength of evidence in this band with $\tau_\mathrm{M}$ averaged also increases by several orders of magnitude compared to that with $\tau_\mathrm{M} = 10^{10}$~yr. In Fig.~\ref{fig:time_avg}, we also find two favored ranges located close to $\log_{10}(\mu/\eV)=-12.10$ and $\log_{10}(\mu/\eV)=-11.75$, with a valley in between. The locations of the favored range on the left (right) is the same as the range in the upper right (left) panel of Fig.~\ref{fg:posterior}. To better understand the dependence of our result on the prior distribution of $\tau_\mathrm{M}$, we show the posterior distributions  with several values of $\tau_\mathrm{M}$ between $10^6$~yr and $10^{10}$~yr in Fig.~\ref{fig:peak_time}. Combining with the two plots in the upper panels of Fig.~\ref{fg:posterior}, we found that the peaks close to $10^{12.08}$~eV shrink and the ones close to $10^{11.80}$~eV rise in height with $\tau_\mathrm{M}$ increasing, but the two favored ranges stay almost unchanged. Therefore, we conclude that the two favored ranges of $\mu$ does not depend on prior distribution of $\tau_\mathrm{M}$, but the their strength of evidence are sensitive to $\tau_\mathrm{M}$.

Since the strength of evidence for the favored mass ranges depends on the prior of the spin and the merger time, it is interesting to ask how large it could be, with the available BH data at the moment. We consider an extreme scenario in which all the BHs have the maximum initial spin, i.e. $\chi=1$. In this case, the role of the superradiant effect for spin-down is maximized. Moreover, we choose two extrema of $\tau_\mathrm{M}$ so that one of the two ranges is the most favored. The results are shown in Fig.~\ref{fig:maximum}. For $\tau_\mathrm{M}=10^6$~yr, we obtain the Bayesian factor $B=73.98$ in the boson mass range $-12.16\leq\log_{10}(\mu/\eV)\leq-12.14$. For $\tau_\mathrm{M}=10^{10}$~yr, we obtain the Bayesian factor $B=95.40$ in the boson mass range $-11.76\leq\log_{10}(\mu/\eV)\leq-11.74$.

\section{Summary and Discussion}\label{sec:summary}

The observed spin distribution of BHs can be used to test the existence of axion-like particles. In this work, we perform a Bayesian analysis of the scalar boson in the mass rage $10^{-13.5}~\eV\leq\mu\leq10^{-11}~\eV$ based on the three phases of GWTC reported by LVK collaboration. The recently improved analytic approximation of the boson cloud eigenfrequency is adapted to calculate the superradiance rate. The Regge trajectories are then obtained from the superradiance rate with the principal number $n$ summed from 0 to 8. The obtained Regge trajectories are very different from previous calculations at the wiggles. We show that the difference cannot be absorbed by redefining the boson mass $\mu$ and the merger time scale $\tau_\mathrm{M}$.

With the new Regge trajectories, the strength of evidence as a function of $\mu$ is sensitive to both the initial spin prior distribution and the merger time scale $\tau_\mathrm{M}$. The value of $\tau_\mathrm{M}$ was fixed in previous studies. In this work, we carefully investigate the effect of varying $\tau_\mathrm{M}$. Two favored ranges of boson mass $\mu$ are found, with one centred at $\log_{10}(\mu/\eV)\sim{-11.75}$ and the other centred at $\log_{10}(\mu/\eV)\sim-12.10$.  Varying the prior distribution of $\tau_\mathrm{M}$ does not change the two ranges, but their strengths of evidence are sensitive to $\tau_M$. Interestingly, the Bayesian analysis with fixed $\tau_\mathrm{M}$ may completely miss one of these two ranges.

We found an exclusion range $-12.80\leq\log_{10}(\mu/\eV)\leq-12.34$ that is not affected by the initial spin prior distribution at fixed $\tau_\mathrm{M}=10^{10}$~yr. This range is consistent with the observation in Ref.~\cite{Ng:2020ruv}, in which $\tau_\mathrm{M}$ is fixed to be $10^7$~yr. Within this range, only $4.8\times10^{-9}$ probability supports the existence of the boson. Nonetheless, the strength of evidence increases by several orders of magnitude if the merger time $\tau_\mathrm{M}$ is averaged. 

We also study the largest strength of evidence could be obtained from the available data, by choosing the extreme scenario in which all BHs are produced with the maximum spin $\chi=1$. With the $\tau_\mathrm{M}$ fixed at $10^6$~yr,  the Bayes factor is 73.98 in the range $-12.16\leq\log_{10}(\mu/\eV)\leq-12.14$. With $\tau_\mathrm{M}$ fixed at $10^{10}$~yr, the Bayesian factor is 95.40 for range $-11.76\leq\log_{10}(\mu/\eV)\leq-11.74$. These results are the upper limits of the evidence strength which can be obtained from the GWTC data by now. We expect better result with more data accumulated.

\begin{acknowledgments}
We thank H. B. Jin for the valuable discussion. This work is supported in part by the National
Nature Science Foundation of China (NSFC) under grants No.~12075136 and the Natural Science Foundation of Shandong Province under grant No.~ZR2020MA094.
\end{acknowledgments}


\end{document}